\documentclass[draft,12pt]{article}
\usepackage{amsmath,amsfonts,palatino,amsthm,epsf}
\newcommand{\beq}{\begin{equation}}
\newcommand{\eeq}{\end{equation}}

\newcommand{\f}{\begin{equation}}
\newcommand{\ff}{\end{equation}}
\newcommand{\blankline}{\vskip .3cm}

\begin{document}

\title{Scientific alternatives to the anthropic principle}
\author{Lee Smolin\thanks{Email address:
lsmolin@perimeterinstitute.ca}\\
\\
\\
Perimeter Institute for Theoretical Physics,\\
35 King Street North, Waterloo, Ontario N2J 2W9, Canada, and \\
Department of Physics, University of Waterloo,\\
Waterloo, Ontario N2L 3G1, Canada\\}
\date{\today}
\maketitle
\vfill
\begin{abstract}

It is  explained in detail why the Anthropic Principle (AP) cannot yield any falsifiable predictions,
and therefore cannot be a part of science.  Cases which have been claimed as successful
predictions from the AP are shown to be not that.  Either they are uncontroversial applications of
selection principles in one universe (as in Dicke's argument), or the predictions made do not
actually logically depend on any
assumption about life or intelligence, but instead depend only on arguments from observed
facts (as in the case of arguments by Hoyle and Weinberg). The Principle of Mediocrity is also examined and
shown to be unreliable, as arguments for factually true conclusions can easily be modified to lead to  false conclusions
by reasonable changes in the specification of the ensemble in which we are assumed to be typical.

We show however that 
it is still possible to make falsifiable predictions from theories of multiverses, if the ensemble predicted has
certain properties specified here.  An example of such a falsifiable multiverse theory is cosmological
natural selection. It is reviewed here and it is argued that the theory remains unfalsified. But it is very
vulnerable to falsification by current observations, which shows that it is a scientific theory. 

The consequences for recent discussions of the AP in the context of  string theory are
discussed.  

\end{abstract}
\vfill

\newpage
\tableofcontents
\newpage

\section{Introduction}

I have chosen a deliberatively provocative title, in order to communicate a sense of
frustration I've felt  for many years about how otherwise sensible people, some of whom 
are among the scientists I most respect and admire, espouse an approach to cosmological problems
that is easily seen to be unscientific.  I am referring of course to the anthropic principle.  
By calling it unscientific I mean something very specific, which is that it fails to have a 
necessary property to be considered a scientific hypothesis. This is that it be 
{\it falsifiable.}  According to  Popper\cite{popper}, a theory is falsifiable if one can derive from it
unambiguous predictions for doable experiments such that, were contrary results seen,
at least one  premise of the theory would have been proven not to apply to nature.  

Having started boldly, I will put the outlines of my argument on the table in a few paragraphs 
here.  The purpose of this essay is then to develop the points in detail.  

 While the notion of falsifiability has been challenged and 
qualified by philosophers since Popper, such as Kuhn, Feyerabend and 
others\footnote{I will not discuss here the history and present status
of the notion of falsifiability, my own views on the methodology of science are discussed
elsewhere\cite{APTS}. } , it
remains the case that  few philosophers of science, and 
few working scientists, would be able to take seriously a proposal for a  fundamental
theory of physics  that had no possibility of being  disproved by a  doable experiment. 

This point is so basic to how science works that it is perhaps worthwhile taking a moment to
review the rationale for it.   Few working scientists will  disagree
that an approach can be considered ``scientific"  only 
to the extent that it requires experts who are
initially in disagreement about the status of a theory to resolve their disagreements-to the fullest
extent possible- by
rational argument from common evidence.    As Popper emphasizes,  science
is the only approach to knowledge whose historical record shows over and over again  
that  consensus was reached among well trained people  as a result
of rational argument from the evidence.   But-and this is Popper's key point- this has only
been possible because proposed theories have been required to be  falsifiable. The 
reason is that the situation is asymmetric: 
confirmation of a prediction of theory does not show that the theory is true, but falsification 
of a prediction can show it is false.  

If a theory is not falsifiable, there is the
very real possibility that experts may find themselves in permanent disagreement
 about it, with no possibility that they
may resolve their disagreement rationally by consideration of evidence.  
The point is that to be part of science, $X$-theorists have to do more than convince other 
$X$-theorists that $X$-theory is true.  They have to convince all the other well trained scientists who
up till now have been skeptical.    If they don't aspire to do this, by rational arguments from the evidence, then
by Popper's definition,  they
are not doing science.  Hence, to prevent the
progress of science from grounding to a halt, which is to say to preserve what makes science
generally successful, scientists have an ethical imperative to consider only falsifiable theories
as possible explanations of natural phenomena.  

There are several versions of the anthropic principle\cite{carrees,carter,barrowtipler}\footnote{My understanding of the 
logical status of the different versions of the Anthropic Principle was much improved
by \cite{sherrie}.}. There is of course the
explicitly theological version, which is by definition outside of science.  I have no reason
to quarrel with that here. I also have no argument against straightforward consideration
of selection effects, so long as the conditions invoked are known independently and
not part of a speculative theory that is otherwise unsupported by any evidence.  I will
discuss this in some detail below, but the short version is that there simply is a vast logical 
difference between taking into account a known fact, such as the fact that most of the
galaxy is empty space, and arguing from a speculative and  unproven premise, such as 
that there is a large ensemble of unseen universes.   

In recent discussions, the version of the anthropic principle that is usually put forward by its
proponents as a scientific idea is based on two premises.

\begin{itemize}

\item{} {\bf A} There exists (in the same sense that our chairs, tables and our universe
exists)  a very large ensemble of  ``universes",  $\cal  M$ which are completely or
almost completely  causally disjoint regions of spacetime, within which the parameters
of the standard models of physics and cosmology differ.   To the extent that they are causally disjoint, we have
no ability to make observations in other universe than our own.   The parameters of the standard models of 
particle physics and 
cosmology vary over the ensemble of universes. 

\item{}{\bf B}  The 
distribution of parameters in $\cal M$ is
random (in some measure) and  the parameters that govern our universe are
rare.  

\end{itemize}

This is the form of the Anthropic Principle most invoked in discussions related to 
inflationary cosmology and string theory, and it is the one I will
critique here.  

Here is the basic argument why a theory based on {\bf A} and {\bf B} is
not falsifiable.  If it at all applies to nature, it follows that our universe
is a member of the ensemble $\cal M$.   Thus, we can assume that 
{\it whatever properties our universe is known to have, or is discovered
to have in the future, it remains true that there is at least one member
of $\cal M$ that has those properties.  Therefore, no experiment, present
or future, could contradict {\bf A} and {\bf B}.}   Moreover, since, by 
{\bf B}, we already assume that there are properties of our universe that
are improbable in $\cal M$, it is impossible to make even a statistical
prediction that, were it not borne out, would contradict {\bf A} and {\bf B} .

There are a number of claims in the literature of predictions made
from {\bf A} and {\bf B}. 
By the logic just outlined, these must all be spurious.  
 In section 5 below I will examine the major claims of this
kind and demonstrate that they are fallacious.   This does not mean that the conclusions
are wrong. As we shall see, there are cases in which the part of the argument that is logically related to the
conclusion has nothing to do with {\bf A} and {\bf B} but instead relies only on
observed facts about the universe. In these cases, the only parts of the argument
that are wrong are the parts that fallaciously attribute the conclusion to a version
of the Anthropic Principle. 

But what if {\bf A} is true?  Will it be possible to do science in such
a universe?  Given what was just said, it is easy to see how a theory
could be constructed so as to still be falsifiable.  To do this
{\bf B} must be replaced by

\begin{itemize}

\item{}{\bf B'}  It is possible nevertheless, to posit a mechanism, $\cal X$ by which the
ensemble $\cal M$ was constructed, on the basis of which one can show that
almost every universe in $\cal M$ has a property $\cal W$, which has the following
characteristics\footnote{As I discuss in section 5  because of the issue of selection effects
connected with the existence of life, this can be weakened to {\it almost every universe in $\cal E$ that
contains life also has property $\cal W$...}.}:

\begin{enumerate}

\item{}$\cal W$ does not follow from any known law of nature or observation, so it is consistent with
everything we know that $\cal W$ could be false in our universe. 

\item{}$\cal P$ There is a doable experiment that could show that $\cal W$ is not true in our
universe.  

\end{enumerate}

\end{itemize}

If these conditions are satisfied then an observation that $\cal W$ is false in our
universe disproves {\bf A} and {\bf B'}.  Since, by assumption, the experiment is
doable, the theory based on these postulates is falsifiable.  

Note that what would be falsified is only the specific {\bf B'} dependent on a 
particular mechanism $\cal X$.   Since $\cal X$ by generating the ensemble,
will imply {\bf A}, what is falsifiable is in fact the postulate that the mechanism
$\cal X$ acts in nature.    Conversely, a mechanism that generates a random
ensemble, as  described by {\bf B} rather than {\bf B'} cannot be falsified, as I will demonstrate
in some detail below. 

Someone might argue that it is logically possible that  {\bf A} and {\bf B} are true and that, if so,  
this would be  bad only for those of us who insist on 
doing science the old fashion way.  If an otherwise attractive
theory points in a direction of {\bf A} and {\bf B} then we should simply
accept this and abandon what may be outmoded ideas about
``how science works".  

If this is the case
then it will always be true that basic questions about the universe cannot
be answered by any scientific theory (that is by a theory that could be rationally
argued, based on shared evidence to be true).  But the fact that is a possibility
does not mean we should worry unduly about it turning out to be true. This is not the only hypothesis
about the world that, if true, means that science must remain forever incomplete.

Similarly,  there are those who argue that it is sufficient to do science with one way
predictions, of the form, 

"Our theory has many solutions, $S_i$.  One of them,  $S_1$ 
gives rise to a prediction $X$.  If $X$ is found that will confirm the combination of our theory and the
particular solution $S_1$.  But if $X$ is not found belief in the theory is not diminished, for there
are a large number of solutions that don't predict $X$."  

One problem with this is that it
can easily lead to a situation in which the scientific community is indefinitely split into groups that
disagree on the likelihood that the theory is true, with no possibility for resolution by rational 
argument.   It is indeed plausible that this is already the case with string theory, which appears
so far unfalsifiable, but can make some claims of this form.  A second problem is that even if 
$X$ were found, it is easy to imagine in this situation another theory could be invented 
that also had a solution that  predicted $X$.  If neither
are falsifiable, there would be no possibility to resolve rationally which were true.

Thus, so long as we prefer a science based on what can be rationally argued from
shared evidence, there is an ethical imperative to examine only  hypothesis
that lead to falsifiable theories. If none are available, our job must be to invent some.  
So long as there are falsifiable-and not yet-
falsified-theories that account for the  phenomena in question, the history of science teaches
us to prefer them to their non-falsifiable rivals.  The simple reason is that once a non-falsifiable
theory is preferred to falsifiable alternatives, the process of science stops and further
increases in knowledge are ruled out.  There are many occasions in the
history of science when this might have happened; we know more than
people who espoused Ptolemaic astronomy, or Lysenko's biology, or Mach and others who dismissed
atoms as forever unobservable, because
at least some scientists preferred to go on examining falsifiable theories. 

Thus, to deflate the temptation to proceed with non-falsifiable theories, it is sufficient
to demonstrate that falsifiable alternatives exist. In this paper I review one falsifiable alternative
to the Anthropic Principle which is cosmological natural 
selection (CNS)\cite{evolution,moreev,lotc}.  As it is falsifiable
it may very well be wrong.

In the last sections of this paper I will review cosmological natural selection in light of developments since it was
first proposed. I will show that, in spite of several claims to the contrary, the theory
has yet to be falsified. However, it remains falsifiable as it makes at least one
prediction for a property $\cal W$ of the kind described in {\bf B'}.    But whether it is right or
wrong, the fact that a  falsifiable theory exists is sufficient  to show that the problems that
motivate the  Anthropic Principle
might be genuinely solved by a falsifiable theory. 

But if the anthropic principle cannot provide a scientific explanation, what are we to make of the claim that the universe is
friendly to life?  It is essential here to   distinguish 
the different versions of the anthropic principle from what I would like to 
call the {\it anthropic observation}.  This observation
states:

\begin{itemize}

\item{}{\bf The anthropic observation:}  {\it  Our universe is much more complex 
 than most universes with the same laws but different values of the parameters of those laws. In particular, 
it has a complex astrophysics, including galaxies and long lived stars, and a complex chemistry, 
including carbon chemistry, These necessary conditions for life are present in our universe as a consequence
of the complexity which is made possible by the special values of the parameters. }

\end{itemize}

I will describe this more specifically below.  There is good evidence that the anthropic observation is
true\cite{carrees,carter, barrowtipler,lotc}.  Why it is true is a puzzle that science must solve.  
To solve it, it does not suffice just to restate
what is to be explained as a principle, especially if the resulting theory that follows from that principle
is  not falsifiable.  One must discover a reason why it is true that has nothing to do
with our own existence.  Cosmological natural selection may be right or it may be wrong, but it does
provide a genuine explanation for the anthropic observation.  This is that the conditions life requires, 
such as carbon chemistry and long lived stars, serve another purpose, in that they contribute to the
reproduction of the universe itself.

\section{The problem of the undetermined parameters of 
physics and cosmology}

The second half of the twentieth century saw a great deal of progress 
in 
our understanding of 
elementary particle physics and cosmology.  In both areas, standard 
models were established, which passed numerous experimental tests.   
In elementary particle physics the standard model, 
described in the mid 1970's is based on two key insights. The first 
is  about unification of the fundamental forces. The second is about 
why that unification does not prevent 
the different particles and forces from have different properties.  
The unifying principle is that all forces 
are described in terms of gauge fields, based on making symmetries 
local. The second principle is
about how the symmetries between particles and among forces can be 
broken naturally when those
gauge fields are coupled to matter  fields.  The standard model of 
cosmology took longer to establish, but
is based also on the behavior of matter fields when the symmetry 
breaks. In particular, this leads generally to the existence of a 
non-zero vacuum energy, which can both drive an early inflation of 
the universe and
act today, accelerating the expansion.

In each case, however there is a catch. The interactions of the gauge 
fields with each other and with
gravity is determined completely by basic symmetries, whose 
description allows a very small number of parameters. However, the 
dynamics of the matter fields needed to realize the idea of the 
symmetry breaking spontaneously and dynamically is arbitrary and 
requires a large number of parameters to describe.
This is because the easiest matter fields to work with are scalar 
fields, and no  transformation
properties constrain the form of their interactions. 

The result is that the standard model of elementary particle physics 
has more 
than 20 adjustable parameters.  These include the masses of all the 
basic stable elementary particles:
proton, neutron, electron, muon, neutrinos etc, as well as the basic 
coupling constants and mixing angles
of the various interactions.   These are not determined by any 
principle or mechanism we know; they must be specified by hand to 
bring the theory into agreement with experiment.  The standard model 
of cosmology
has similarly about fifteen parameters.  

Two of the biggest mysteries of modern science are then how these 35 
or so parameters are determined.

There are two especially puzzling aspects to these problems

The first is the {\it naturality problem}. Many of these parameters, 
when expressed in terms of dimensionless ratios, are extremely tiny 
or extremely large numbers. In Planck units, the proton and neutron 
masses are around $10^{-19}$, the cosmological constant is 
$10^{-120}$, the coupling constant for
self-interactions of the field responsible for inflation cannot be 
larger than  $10^{-11}$ and so on. 

The second is the {\it complexity problem.}  Our universe has an 
array of complex and non-equilibrium structures spread out over a 
huge range of scales from clusters of galaxies to living cells. It is 
not too hard to see that this remarkable circumstance depends on the 
parameters being fine tuned into narrow windows.
Were the neutron heavier by only one percent, the proton light by the 
same amount, the electron twice as massive, its electric charge 
twenty percent stronger, the neutrino as massive as the electron 
etc.  there would be no stable nuclei at all.  There would be no 
stars, no chemistry.  The universe would be just hydrogen gas. 
The anthropic observation stated in the introduction is one way to state the complexity problem. 

Despite all the progress in gauge theories, quantum gravity, string 
theory etc. not one of these problems have been solved.  Not one mass 
or coupling constant of any  particle considered now to be elementary 
has ever been explained by fundamental theory.  

\section{The failure of unification to solve the problem}

For many decades there has been a consensus on how to solve the 
problems of the undetermined 
parameters: {\it unify the different forces and particles by increasing the symmetry of the theory and the 
number of parameters will decrease.}
The expectation that unification reduces the number of parameters in 
a theory is due both to historical experience and to a philosophical 
argument.  The former is easy to understand: It worked when 
Newton unified the theory of the planetary orbits. It worked when 
Maxwell showed that light was a consequence of the unification of 
electricity and magnetism and so on.  In several cases, unification 
was accomplished by the discovery of a symmetry principle, and by 
relating things heretofore unrelated, a new symmetry principle can 
sometimes reduce the number of parameters.  The philosophical 
argument  is along the lines of the following: 
reductionism will lead to a fundamental theory, a fundamental theory 
will answer all possible questions and so can't have free parameters, 
and unification operates in the service of greater reductionism.    
Or perhaps: the theory that unifies everything should be able to 
answer all questions. So it had better be
unique, otherwise there would be unanswerable questions, having to do 
with choosing which
unified theory corresponds to nature. 

Whatever the arguments for it, the correlation between unification 
and reduction in the number of parameters
has not worked recently.  Indeed, the last few times it was tried, it 
went the other way. 
 One can reduce by a few the number of parameters 
by unifying all of elementary particle physics in one Grand Unified 
Theory. One does not eliminate most of the freedom, because the 
masses of all the observed fundamental parameters are traded for the 
values of Higgs vacuum expectation values, which are not determined 
by any symmetry and remain
free parameters.  

The grand unified theories had two problems. The first is that the simplest
version of them, based on the group $SU(5)$, was falsified. It predicted that
protons would decay, with at least a certain rate. The experiment was done, and
protons were seen not to decay at that rate.  This is the last time there was a significant
experimental test of a new theoretical idea about the elementary particles. 

One can consider more complicated grand unified models in which protons
don't decay, or the decay rate is much smaller. But all such models suffer
from the second problem. This is the problem of naturality. They 
require two Higgs scales, one at around $1 Tev$ 
and one at about $10^{15} Tev$'s.  But quantum corrections tend
to pull the two scales closer to each other, To keep their ratio so 
large 
requires fine tuning of the coupling constants of the theory to 
roughly a part in the ratio or
$10^{-15}$.   

To solve this problem, it has been proposed to supersymmetrize the 
model.  Supersymmetry
relates bosons to fermions, and one might think  it reduces the 
number of free parameters,
but it goes the other way. The {\it simplest}  
supersymmetric extension of the  standard model has
more than $100$ parameters.

Supersymmetry is a beautiful idea, and it was hard not to get very excited about it when it
was first introduced.  But so far it has to be counted as a disappointment.  Had the
addition of supersymmetry to what we know led to unique predictions, for example
for what will be seen at the $LHC$, that would have been very compelling. The 
reality has turned out to be quite different.  
The problem is that, while supersymmetry is not precisely unfalsifiable, it is difficult to falsify, as many 
negative results can be-and have been-
dealt with by changing the parameters of the theory.   Supersymmetry would be 
completely convincing if there were even  one pair, out of all
the observed fundamental particles,  that could be made into superpartners.  
Unfortunately,  this is not the case and 
one has to invent superpartners for each one of the presently observed
particles.  

This turns out to introduce a huge amount of arbitrariness.  
The current situation is that the minimal super symmetric standard model has so much
freedom, coming from its $105$ dimensional parameter space, that, depending on which 
region of the parameters space one chooses,  there are at least a
dozen scenarios for what could be seen by the upcoming $LHC$ experiments\cite{superfreedom}.  It is not
too much of an exaggeration to say that almost any results  that could be seen by the $LHC$
could be-and probably will be-promoted as evidence for supersymmetry, whether or not
it actually is.  To actually test whether or not particle physics is supersymmetric will take much longer as it will 
require measuring enough amplitudes to see if they are related by supersymmetry.  

Another possible solution is to further unify the theory, by coupling to 
gravity.  However, this seems either to not help, or to make things 
much worse.  There are two well developed approaches to
quantum gravity, one non-perturbative, which means it makes no use of 
a background classical
spacetime and one perturbative, which describes small excitations of 
a classical spacetime, 
The non-perturbative approach, called {\it loop quantum gravity}, 
readily incorporates coupling to the standard model of elementary 
particle physics and also readily incorporates supersymmetry.
It does not seem that, from a non-perturbative point of view, 
coupling to quantum gravity constrains
the parameters of particle physics.

The perturbative approach, which is string theory, makes very strong 
assumptions about how the string is to be quantized\footnote{A recent 
paper by Thiemann\cite{thomas-string} suggests that with different 
technical assumptions there are consistent string theories in any 
dimension, without supersymmetry.} and it also makes two physical 
assumptions: 1)  that no matter how small one looks, spacetime looks 
classical, with small quantum excitations 2) Lorentz invariance is a 
good symmetry out to infinite energies and boosts.   It is not 
certain that all these assumptions can be realized consistently, as 
after many years there are only proofs of consistency and finiteness
of perturbative string theory to second non-trivial order in 
perturbation theory, and attempts to go further have not so far 
succeeded\footnote{For details of precisely
what has and has not been proven regarding string theory, loop quantum gravity and 
other approaches to quantum gravity see\cite{wherearewe}} .  But these results indicate that the assumptions just 
mentioned do put some constraints on particle physics. Supersymmetry 
is required and the dimension of spacetime must be ten.  

To the order of perturbation theory it is known to be consistent, 
string theory unifies all the interactions, including gravity.  It 
was therefore originally hoped that it would be unique. These hopes 
were quickly bashed, and indeed, the number of string theories for 
which there is some
evidence for has been growing exponentially as string theorists 
developed better techniques
to construct them.  Originally there were five consistent 
supersymmetric string theories in ten
dimensions. But, of course, the number of observed dimensions is 
four. This led to the hypothesis that the extra six dimensions are 
curled up in small spaces, or otherwise hidden from
large scale observations.  Unfortunately, the number of ways to do 
this is quite large, at
least $10^5$.  In recent years evidence has been found for many more 
string theories,
which incorporate non-perturbative structures of various 
dimensions, called branes.  

A key problem has been constructing string theories that agree with 
the astronomical
evidence that the vacuum energy (or cosmological constant) is 
positive.  The problem
is that a positive cosmological constant is not consistent with 
supersymmetry. But supersymmetry
appears to be necessary to cancel dramatic instabilities having to do 
with the existence
of tachyons in the spectrum of string theories. 

A year ago dramatic progress was made on this problem by Kachru\cite{kachru} and 
collaborators\footnote{building on earlier work by Giddings et al, \cite{Giddings}, 
Bousso and Polchinski\cite{BP} and others.}.
They found a way to sneak up on the problem by wrapping branes 
around cycles of
the compactified six manifold.  They discovered evidence for the 
existence of string theories
with positive vacuum energy.  This evidence is very weak-they are 
unable for example to
construct propagation amplitudes even for the free, non-interacting 
strings.  What they are
able to do is to argue that if there are consistent string theories 
with the desired characteristics,
their low energy behavior should be captured by solutions to 
classical supergravity, coupled to
the patterns of branes in question.  Then they construct the low 
energy, classical, supergravity
description.  

Of course the logic here is backwards. Had they been able to show 
that the required supergravity solutions don't exist, they would have 
ruled out the corresponding string theories.  But the existence of a 
good low energy limit is a necessary but not sufficient condition for 
a theory to be shown to exist.  So on logical grounds, the evidence 
for string theories with positive cosmological
constant is very weak.

However, if one takes the possibility of the existence of these 
theories seriously, there is a 
disturbing consequence. For the number of distinct  theories that 
the evidence points to is vast,
estimates have been made on the order of $10^{100}$ to $10^{500}$ \cite{Susskind,Douglas}.   
Each of these
theories is consistent with the macroscopic world being four 
dimensional, and the existence
of a positive and small vacuum energy. But they disagree about 
everything else, in particular,
they imply different versions of elementary particle physics, with 
different gauge groups, spectra
of fermions and scalars and different parameters.  

That is, if the string theorists are right, there are on the order of 
$10^{100}$ or more different
ways to consistently unify gauge fields, fermions and gravity. {\it  This 
makes it likely\footnote{It should be noted that while some string theorists have argued that
this situation calls for some version of the Anthropic Principle\cite{Susskind}, other have sought
ways to still pull falsifiable predictions from the theory\cite{Douglas}.}
that string theory will never make any new, testable predictions
concerning the elementary
particles}\footnote{Of course, it may be correctly claimed that string theory makes
a small number of correct {\it postdictions}, for example that there are fermions, gauge fields and
gravitational fields in nature, and 
that there are no more than $10$ spacetime dimensions.   But this does not by itself
give a strong argument for the truth of string theory as there are other approaches
to unifying gravity with quantum theory and the standard model about which
non-trivial properties have also been proven\cite{wherearewe}.  So there is no evidence
that string theory is the unique theory that unifies gravity with the standard model. }.
  
Of course, a very small proportion of the theories will 
be consistent with the
data we have, to date, about particle physics. Suppose this is only 
one in
$10^{50}$.  There will still be $10^{50}$ different theories, which 
will differ on
what we will see in future experiments at higher energy.  This number 
is so vast,
it appears likely that whatever is found, there will be many versions 
of string theory that
agrees with it.

\section{Mechanisms of production of universes}

Whatever the fate of the positive vacuum energy solutions of string 
theory, one thing is
clear. At least up till now, the hope that unification would lead to 
a unique theory has
failed, and it has failed dramatically.  So it seems unlikely that 
the problem of accounting for 
the values of the parameters of the standard models of particle 
physics and cosmology
will be solved by restrictions coming from the consistency of a 
unified theory.

The rest of this essay is then devoted to alternative explanations of 
the choice of parameters.
All alternatives I am aware of involve the postulate {\bf A}.  They
also require 

\begin{itemize}

\item{}{\bf C}  There are many different 
possible consistent  phenomenological descriptions  of physics, 
relevant for the possible description of elementary particle physics 
at scales much less than Planck energies.  These may correspond to 
different phases of the vacuum, or different theories altogether. 

\end{itemize}

As a result, fundamental physics is assumed to give us, not a single theory,
but a space of theories, $\cal L$, which has been called 
the {\it landscape} \cite{moreev,lotc}\footnote{It is perhaps worth mentioning that
the word ``landscape" was chosen  in \cite{moreev,lotc} to make the transition to the 
concept of {\it fitness landscape}, well known in evolutionary theory, more transparent.}.
 As in biology, we distinguish the space of genotypes from the space of phenotypes
will have to distinguish $\cal L$ from the space of the parameters of the standard
model, which  we will call $\cal P$.   

All multiverse theories  then make some version of the 

\begin{itemize}

\item{}{\bf Multiverse hypothesis.}  Assuming {\bf A} and {\bf C}, the    
whole of reality-which we call the multiverse- consists of many different 
regions of spacetime, within
which phenomena are governed by different of these phenomenological 
descriptions.  For simplicity, we
call these {\it universes}.    
\end{itemize}

The  multiverse is then described by probability distributions $\rho_L$ in $\cal L$
and $\rho_P$ in $\cal P$.  These describe the population of universes within the
multiverse. 
Multiverse theories can be classified by the answers to three questions.

\begin{enumerate}

\item{}How is the ensemble of universes generated?

\item{}What is the mechanism that produces the probability distribution
$\rho_P$?

\item{}What methodology is used to produce predictions for our universe
from the ensemble of universes?

\end{enumerate}

We will be 
interested here only in those multiverse theories that make 
falsifiable predictions. To do this the ensemble of universes cannot 
be arbitrarily specified, otherwise it could be arbitrarily adjusted 
to agree with any observations for example by making a typical 
universe agree
with whatever is observed about ours.  To have empirical content, the 
ensemble of universes must be
generated by some dynamical mechanism and that mechanism, in turn, 
must be one that is a consequence
of general laws.  That way, the properties of the ensemble are 
determined by laws that have other
consequences, and, at least in principle, can be checked, 
independently.

Two mechanisms for generation of universes have been studied, eternal 
inflation and bouncing
black hole singularities.  We describe each of them, after which we 
will contrast their properties.

\subsection{Eternal inflation}

The hypothesis of early universe inflation gives a plausible 
explanation of several observed features
of our universe, such as its homogeneity and uniformity\cite{basicinflation,newinflation}. The basic 
idea is that at very early times the 
energy density is dominated by a large vacuum energy, possibly coming 
from the vacuum 
expectation value of a scalar field.  As the universe expands 
exponentially, driven by the vacuum energy,
the vacuum expectation value also evolves in its potential.  
Inflation comes to an end when a local
minimum of the potential is reached, converting vacuum energy into 
thermal energy that is presumed
to provide the thermal energy that becomes the observed cosmic 
microwave background.  

The model appears to be consistent, assuming all scales involved are 
less than the Planck scale, and has made predictions which were 
confirmed\footnote{But it should be noted that other theories make 
predictions
so far indistinguishable from those of inflation\cite{pauland,VSL}.}. But there 
are some open problems. One set has to do with the initial conditions 
necessary for inflation to happen.  It has been shown that a region 
of spacetime will begin to inflate if the vacuum energy dominates 
other sources over its extent and, the matter and gravitational 
fields are
constant to good approximation over that region.   Of course, we do 
not know the initial conditions for the universe,
and we observe nothing so far about the conditions prior to 
inflation.  But on several plausible hypotheses about
the initial state, the conditions required for a region of spacetime 
to begin inflating are improbable.  For example, the existence of inflation,
together with the smallness of $\delta \rho /\rho $, requires that the self-coupling
of the inflaton be small.  

However, once the conditions necessary for inflation are met, it appears to be, in
some models, likely that inflation does not happen just once.    The reason is that
because of quantum fluctuations, the scalar field will sometimes fluctuate ``up"
in the potential. The result is that even after inflation has ended in one region, it
will continue in other regions. This can lead to the scenario known as
{\it eternal inflation}\cite{alex-eternal,linde-eternal} in which there are always regions which continue to
inflate. There is then a competition between the classical force from the potential,
causing the expectation value to decrease or ``roll" towards a local minimum, and the
quantum fluctuations which can lead it to locally increase. Given plausible, but not
necessary assumptions\cite{alex-eternal,linde-eternal}, this can result in the creation
of a large, or even infinite, number of regions which locally resemble ordinary 
$FRW$ universes.

\subsection{Bouncing black hole singularities}

A second mechanism for generating new universes is through 
the formation of black holes. It is known that a collapsing
star, such as a remnant of a supernova, will form either
a neutron star or a black hole, depending on the mass. There
is an upper mass limit (UML) for a stable neutron star, remnants
of supernovae over this mass have nothing to restrain them and will
collapse to the point that a horizon is formed.  We have rough
estimates for the value of the upper mass limit, between
$1.5$ and $2.5$ solar masses. 

According to the
singularity theorem of Penrose, proved on general assumptions, 
classical general relativity predicts that a singularity will
necessarily form, at which the curvature of spacetime becomes
infinite and spacetime ends.  No trajectory of a particle or photon can'
be continued passed the singularity to the future.  

However, this result may be modified by quantum effects. Before the 
singularity is reached, densities and curvatures become Planck scale
and the right dynamics will by those of the quantum theory of 
gravity. As early as the 1960's pioneers of the field of quantum
gravity such as John Wheeler and Bryce deWitt conjectured that 
the effects of quantum gravity would reverse the collapse, removing 
the singularity
and causing the matter that was collapsing to now expand\cite{old}.  
Time then does not end, and there is a region of 
spacetime to the future of where the singularity would have been.  
The result is the creation of a new expanding region of spacetime, 
which may grow and become for all practical purposes a new
universe. This region is inaccessible from the region where
the black hole originally formed. The horizon is still there, which
means that no light can escape from the new region to the previous 
universe.  From the point of view of causal structure, unless
the black hole evaporates, every event in the new region is to the
future of every event in the region of spacetime where the black hole
formed. 

The transition by which  a collape to a singularitiy is replaced by 
a new expanding region
of spacetime is called a bounce. One can then hypothesize that our
own big bang is the outcome of a collapse in a previous universe and
that every black hole in our universe is giving rise to a new
universe.  

The conjecture that singularities in
classical general relativity are replaced by bounces has been 
investigated and confirmed in many semiclassical calculations\cite{bounce}.
It is also suggested by some calculations in string theory\cite{string-bounce}.    
In recent years the quantum theory of gravity has been developed
to the point where the conjecture can be investigated exactly.
It has been shown that cosmological singularities do bounce\cite{bojowald}.
This means that, assuming that the quantum theory of gravity is 
correct, the big bang in our past could not have been the first
moment of time, there must have been something before that. Results of
the same reliability have yet to be published for black hole 
singularities, but they are in progress\cite{bojowald-personal}.

\subsection{Comparison of universe generation mechanisms}

It is useful to compare the two mechanisms of reproduction of 
universes.  

\begin{itemize}
   
    \item{}{\it How reliably is the evidence for the mode of
    production of universes?}

    We know that our universe contains black holes. 
    There is observational evidence 
    that many galaxies have large black holes in their
    centers.   Black holes are also believed to form from
    supernova remnants, and there is believed to be 
    around $10^{18}$ such black holes.   A number of candidates for such stellar-mass
black holes have been found, and the evidence so far, for example that coming
from studies of X-rays from their accretion disks, supports their identification as
black holes.  
    
    There have been speculations that
    many black holes may have been created by strong inhomogeneities
    in the early universe. However, theories of inflation predict that
    inhomogeneities were not strong enough to create many such 
    primordial black holes. In any case, were they there, one would 
    expect to see signals of their final evaporation, and no such 
    signals have been detected. Thus, it is likely that the population
    of black holes is dominated, numerically, by supernova remnants. 
    
    We also have reasonable, if not yet compelling, theoretical evidence that black holes
    bounce\cite{bounce,string-bounce}. 
And we have exact results that show that quantum effects
    remove the singularity to our past, implying that there was 
    something to the past of our big bang\cite{bojowald}.  
    
Thus, there is plausible  evidence that our universe is creating new
universes through the mechanism of black hole production, and 
that our own universe was created by such a process.  
    
By contrast, the process for formation of new universe in eternal 
 inflation cannot be observed, for it takes place outside of our 
 past horizon.  The existence of the process depends entirely on 
 believing in particular inflationary models that lead also to eternal inflation.  
While many do, it is also possible to invent inflationary models that do not
lead to eternal inflation.  While there is evidence for
 inflation in general, as several predictions of inflation have
  been confirmed by observations,  the observations do not distinguish
so far between different versions of inflationary models, and so cannot
distinguish between models that do and do not predict eternal inflation.  

It is also the case that some, but not all, of the calculations backing up the eternal 
inflation scenario are done using very rough methods, based on
imprecise theories employing semiclassical estimates for
 ``the wavefunction of the universe". This is a speculative extension of
quantum theory to cosmology which has not been put on firm ground,
conceptually or mathematically.  Very recently, progress has been made in
quantum cosmology which does allow precise predictions to be made from a 
rigorous framework\cite{bojowald}. However, while inflation has been studied
with these methods, so far the results do not address the conjectures that
underlie eternal inflation.  

Other approaches to eternal inflation\cite{alex-eternal} rely only on quantum
field theory in curved spacetime.  This is better understood, but there are still
open questions about its applicability for cosmological questions.  
    
    As a consequence, eternal inflation can be considered an 
    interesting speculation, but it is supported neither by observation 
    nor by firm mathematical results within a well defined theory of quantum
gravity.  
    
    \item{}{\it What physics is involved in the mechanism of
    reproduction of universes?  How well do we understand the
    processes that govern the numbers of universes which are created?}
    
    The physical scale governing the birth of universes in eternal 
    inflation is the scale of the inflaton potential in the regime
    where nucleation of new inflating regions take place. This is
at least the grand unified scale, ($\sim 10^{15} Gev$) and
    could be up at the Planck scale. We have theories about
    the physics at this scale, but so far no predictions made by
    these theories, have been confirmed 
    experimentally.  In fact, the only experimental evidence we have concerning
this scale, coming from proton decay experiments, falsified the simplest
grand unified theories. 
    
    By contrast, the physical scale that governs black hole production
    is that of ordinary physics and chemistry. How many stars massive 
    enough to supernova is determined by ordinary chemical processes 
    that govern the formation and cooling of giant molecular clouds.
    We know the physics of stars and supernovas reasonably well, and
    knowledge is improving all the time due to progress in theory,
    observation and experiment.  
    
    Thus, the answer is that we understand well the physics that 
    controls how many universes are created through black hole 
    formation, while we have speculations, but no confirmed or detailed
    understanding of the processes that govern the creation of new
    universes in eternal inflation.  
    
    \item{}{\it What is the structure of the multiverse predicted 
    by each theory?}
    
    A multiverse formed by black holes bouncing looks like a family 
    tree. Each universe has an ancestor, which is another universe.
    Our universe has at least $10^{18}$ children, if they are like
    ours they have each roughly the same number of their own.
    
    By contrast, the structure of a multiverse formed by eternal 
    inflation is much simpler.  Each universe has the same ancestor,
    which is the primordial vacuum.  Universes themselves have no
    descendents.

\end{itemize}

\section{How do multiverse theories make predictions?}

Just as there are two modes of production of universes, there are
two modes of explanation by which people have tried to draw physical
predictions from multiverse models. These are the Anthropic Principle
(AP) and Cosmological Natural Selection (CNS). 

\subsection{Varieties of anthropic reasoning}

There are actually several different anthropic principles
and several different ways that people reason from them to conclusions.
We discuss the major ones here, including the examples that are
usually cited as successes of the Anthropic Principle, which are the
arguments of Dicke, Hoyle and Weinberg\footnote{Note that I do not use the traditional nomenclature
of weak and strong anthropic principles, as these have been used
to refer to different ideas in different books and papers. }.

\subsubsection{The theological anthropic principle}

It is not surprising that some theologians and scientists take
the {\it complexity} problem as evidence that our universe was created
by a benevolent God. They argue that if the best efforts of science 
lead to an understanding of laws of nature within which there is 
choice, and if the choices that lead to a universe with intelligent
life are extremely improbable, the very fact that such an improbable
choice was made is evidence for intension.  This is of course the old
argument from design, recycled from controversies over evolutionary theory.
It should be admitted that it does have force: the discovery of a craft
as complex as an airbus on a new planet would be good evidence for
intelligent life there. But this argument 
has force  only so long as there are no plausible alternative
explanations for how the choice might have been made.  In the case
of biology, natural selection provides a falsifiable and so far 
successful explanation, which renders unnecessary the argument from 
design. 

We can learn from the long history of the controversy in biology what
tests a proposed explanation must satisfy if it is to be more 
convincing than the argument from design for intentional creation of a 
life friendly universe.

\begin{enumerate}
    
    \item{}There must be a physical mechanism which converts the 
    improbable to the probable, that is that raises the probability
    that a universe like ours was chosen from infinitesimal to order
    unity. 
    
    \item{} That mechanism must be falsifiable. It must be built from
    processes or components which can be examined empirically and
    be seen to function as hypothesized, either by being created in a'
    laboratory or by occurring in nature in our observable universe. 
    
\end{enumerate}    

We will see below that these are not satisfied by the different
versions of the Anthropic Principle used in physics and cosmology.
After this we will see a way to reason about multiunverses that is not anthropic
that does satisfy these two tests. 

\subsubsection{Selection effects within one universe}

The first arguments called ``anthropic" in cosmology were based on
the use of {\it selection effects} within our observable universe. 
A selection effect is an effect due to the conditions of observation,
which must be applied to a set of observations before they can be
interpreted properly.  A classical example is the following: Early 
human beings observed that all around them was land and water, and 
over them is sky. From 
this they perhaps deduced that the universe consists of a
vast continent of land, surrounded by water, over which is sky.  
They were wrong, and the reason was that they forgot to take into 
account that the conditions they observed were necessary for them
to exist, as intelligent mammals. We now know that the universe is 
vastly bigger than they imagined and most of it is filled with nothing but a very dilute gas and 
radiation.  If we picked a point randomly, it would be very unlikely 
to be on the surface of a planet. But the conditions necessary for 
our evolution turn  the improbable into the probable. 

Dicke used this logic to debunk a claim of Dirac, concerning his
``law of large numbers". Dirac observed a coincidence
between the age of the universe in Planck units and 
the proton mass in Planck units (the first is roughly
the inverse cube of the second). He argued that this required explanation, and
proposed one, according to which Planck's constant would change in 
time.  Dicke pointed out that the coincidence could be explained 
entirely by our own existence. Intelligent life requires billions of years of
evolution on the surface of a planet near a stable long lived star,
and he was able to argue that the physics of stars imply that 
these conditions would only hold at an era in the universe where Dirac's 
coincidence would hold.  Indeed, when checked, Dirac's proposed 
explanation was falsified. 

This argument is, so far as I know, logically sound. But notice as we go 
along that it is logically quite different than the proposed modes of 
explanation we are about to discuss. The fact that this argument
is sound is not evidence for arguments that have a very different 
logical structure and empirical grounding.

\subsubsection{False uses of an anthropic principle}

There are other successful arguments, which have been called 
``anthropic", although they have nothing to do with selection
effects or the existence of life.  An illustrative example is Hoyle's
prediction of a certain resonance in the nuclei of carbon.

Hoyle argued that for life to exist there must be carbon. 
Indeed, carbon is plentiful in the universe. It must have been made sometime in the history
of the universe, either during the big bang, or afterwards, in stars, as these are the only
ways the universe synthesizes copious amounts of chemical elements.  Detailed studies
show it could not have been made in the big bang, hence we know carbon must have
been made in stars.   
Hoyle was able 
to argue that carbon could only be formed in stars if there were 
a certain resonant state of carbon nuclei. He communicated this 
prediction to a group of experimentalists who found it.  

The success of Hoyle's prediction is sometimes used as support for
the effectiveness of the anthropic principle. But notice that it has
nothing whatsoever to do with the existence of life or intelligence.
The first line of the last paragraph has no logical relation to the
rest of the paragraph.  The logic  of  Hoyle's argument starts with the assertion
of the  empirically well
established fact that the universe is full of carbon.  He then reasons correctly
from this that there must be a source of carbon. Since calculations 
show it cannot be made in the big bang, the only plausible site of 
carbon production is stars. Hoyle analyzed fusion processes in stars
and concluded that carbon would not be produced unless that particular
resonance existed.  

Thus, Hoyle's argument is sound, and led to a successful prediction.
But the first line of his argument is unnecessary. The fact that we, 
or other living things are made of carbon is totally unnecessary to the
argument, indeed were there intelligent life forms which evolved without 
carbon chemistry, they could just as easily make Hoyle's argument. 

To be clear about why Hoyle's argument does not employ any version of the Anthropic
principle, let us examine its logical schema and then ask which step we would have to
question were the prediction falsified.

{\bf Hoyle's argument:}

\begin{enumerate}

\item{}$X$ is necessary for life to exist.

\item{}In fact $X$ is true about our universe.

\item{}Using the laws of physics, as presently understood, together with
perhaps other observed facts, $Y$, we deduce
that if $X$ is true of our universe so is $Z$.

\item{}We therefore predict that $Z$ is true.

\end{enumerate}

In Hoyle's case, $X$ is that the universe is full of carbon, $Y$ is the claim that it could
only be made in stars, and $Z$ is the existence of a certain resonance in carbon.

We see clearly that the prediction of $Z$ in no way depends on step 1.  The argument has the same
force if step 1 is removed. To see this ask what we would do were $Z$ found not to be true.
Our only option would be to question either $Y$ or the deduction from the presently known laws
of physics to $Z$. We might conclude that the deduction was wrong, for example if we made a mistake
in a calculation. If no such option worked, we might have to conclude that the laws of physics might
have to be modified.  But we would never question 1, because, while a true fact, it plays no
role in the logic of the argument leading to the prediction for $Z$.

There are other examples of this kind of mistaken reasoning, in which an argument
promoted as ``anthropic" actually has nothing to do with the existence of
life, but is instead a straightforward deduction from observed facts.  
We will see below  that one famous
argument of Weinberg concerning the cosmological constant is of this kind.

\subsubsection{Selection effects within a multiverse}

More recently, arguments called ``anthropic" are made within 
multi-universe scenarios.  It is a bit tricky to pull falsifiable
predictions from such a scenario, for the simple reason that we
only observe (so far) only one member of the ensemble. But, it is
not impossible, as I will shortly show. 

First, however, we have to dispose of mistaken uses of multi-universe selection effects. 
These are arguments in which point 1, in the schema for Hoyle's argument, is
replaced by

\blankline

{\bf 1'} We live in one member of a multiverse in which the laws of physics vary.
$X$ is necessary for life, therefore by a selection effect we must live in a universe
in which $X$ is true.

\blankline

The full argument is now

\blankline

{\bf Multiverse version of Hoyle's argument:}

\begin{enumerate}

\item{}{\bf'} We live in one member of a multiverse in which the laws of physics vary.
$X$ is necessary for life, therefore by a selection effect we must live in a universe
in which $X$ is true.

\item{}In fact $X$ is true about our universe.

\item{}Using the laws of physics, as presently understood, together with
perhaps other observed facts, $Y$, we deduce
that if $X$ is true of our universe so is $Z$.

\item{}We therefore predict that $Z$ is true.

\end{enumerate}

We see the substitution of {\bf 1'} for {\bf 1} has not changed the logic of the argument.  {\bf 1'} is as 
irrelevant for the argument as {\bf 1} was, because {\bf 2} still does the real logical work.  Furthermore,
if in this case the prediction $Z$ were falsified, we would certainly not question
{\bf 1'}.  The problem we would then have is in understanding why, {\it in one universe}, 
$X$ is true without $Z$ being true.   The problem must be solved within one universe, it
is entirely irrelevant for making the absence of $Z$ consistent with $X$ whether or not the
universe we live in is the only universe or part of a multiverse.  

That is, had the resonance lines of carbon which Holye predicted not been found, 
Hoyle would not have questioned  the existence of life.  Neither would he have 
thought the result relevant to the question of how many universes there are. Instead,
given that carbon is plentiful, he would have examined all the steps in the argument, looking for
a loophole, till he found one.  The loophole would have been something like, there are other sources of 
carbon production, not yet known, for example, in exotic events such as collisions of neutron stars. 

Hence, to pull a genuinely
falsifiable prediction from a multiverse theory,  that genuinely depends logically on the hypothesis that our
universe is part of a multiverse, the logic
must be   different from  the schema just given. 

Here is one way to do it.  Let us fix a multiverse theory, called 
$\cal T$. This theory gives rise to an ensemble of universes, 
$\cal M$.  We are interested in predictions concerning 
some properties $p_i$, where $i$ labels a set of possible properties.
The theory $\cal T$ may give us some {\it a priori}  probability
$\rho_{\cal M} (i)$ that if a universe is picked randomly from the
ensemble, $\cal M$,  it will have property $i$. 

To make the argument below precise, we will need to refer to
another ensemble, which is an ensemble of randomly generated
universes, $\cal R$. This is produced by taking properties allowed
to vary within the theory, and selecting their values randomly, according to 
some measure on the parameter space of the theory.  By random we mean 
that the measure chosen is unbiased with respect to choice of hypothesis as to
the physical mechanism that might have produced the ensemble.
For example, if we are interested in the string vacua, we simply
pick randomly universes with different string vacua. The difference
between $\cal R$ and $\cal M$ is that the former is picked randomly
from the physically possible universes, whereas the latter is 
generated dynamically, by a mechanism proscribed in the theory,
$\cal T$.  

But before comparing this with our universe we should take into 
account that we may not live in a typical member of the ensemble
$\cal M$. There will be a sub-ensemble ${\cal LM} \subset {\cal M}$
of universes that have the conditions for intelligent life to exist.
Depending on the theory, the probability for a random universe in
$\cal M$ to be also in $\cal LM$ may be very small, or it may be close 
to unity.  But we already know that, if the theory is true, 
we are in a universe in $\cal LM$.
So we should compute $\rho_{\cal L} (i)$ the property that a universe
randomly picked in the sub-ensemble $\cal L$ contains property 
$p_i$.

Similarly, there will be a sub-ensemble ${\cal LR} \subset {\cal R}$
of those universes within the random ensemble, which contain life. 

The theory then can only make a falsifiable prediction if some 
restrictive
conditions are satisfied. It is no good to consider properties
that depend on the conditions necessary for life, for they will
always be satisfied in a universe where life exists. 
To find a falsifiable prediction, the following must hold:

I) There must be a 
property $\cal B$ which is
logically independent of the existence of life, that is it must
be physically and logically possible that universes exist which
have life but do not have $\cal B$. To make this meaningful, we
must refer to the ensemble $\cal R$ of random universes. It must
be the case that the probability for a universe in $\cal LR$ to have
property $\cal B$, is small.

II) It must nevertheless be true that within the ensemble 
$\cal M$ generated by the theory $\cal T$ there is strong correlation
between universes with life and universes with property $\cal B$.

III) The argument will have force if the property $\cal B$ has
not yet been looked for, so that the prediction of $\cal B$
is a genuine prediction of the theory $\cal T$, one which is
vulnerable to falsification at the time the prediction is made. 

Under these conditions we can now proceed to do real science
with a multiverse theory. We make the assumption that our universe is a 
typical member of the ensemble $\cal L$. We then look for
property $\cal B$. The theory is falsifiable because if 
property $\cal B$ is not seen in our universe than we know that
theory $\cal T$ that gave rise to the ensemble $\cal L$ is
false.  If, however, $\cal B$ is found, then the evidence favors
the ensemble $\cal M$ produced by the theory over the 
random ensemble $\cal R$.

We can draw a very important conclusion from this. To make a 
falsifiable prediction, a theory must produce an ensemble
$\cal M$ that differs from a random ensemble, $\cal R$, generated
by choosing physically possible universes randomly. There must
be properties that are improbable in $\cal LR$ and
probable in $\cal LM$.  Why is this so? If the two ensembles
are identical, then if there is a high probability that a universe
with life in $\cal M$ has property $\cal B$, this is also true
of a universe with life in a randomly generated ensemble. There are
two problems with this. First, 
the particular hypotheses that make up the theory $\cal T$ are not
being tested, for they are empirically equivalent to a random number
generator.  Second, and more importantly, without the random ensemble,
we cannot give meaning to the necessary condition that $\cal B$
is uncorrelated with the conditions necessary for life. For the
observation of $\cal B$ to have the possibility of falsifying the
theory $\cal T$, it must be logically possible that there exist
ensembles in which the probability of $\cal B$ in universes with life
is low. The operational meaning of this is that they are uncorrelated
in an ensemble of randomly generated universes. 

To put this more strongly, suppose that a theory $\cal T^\prime$
generates an ensemble whose living sub-ensemble $\cal LM$ is
identical to the living sub-ensemble of the random ensemble.
Supposing that theory predicts a property $\cal B$ because
it has probability close to unity in $\cal LM$. If $\cal B$ is
observed, however, that does not stand as evidence for
$\cal T$, because  there is already a complete correlation
between $\cal B$ and life in the ensemble of randomly generated
universes.  

The conclusion is that no multiverse theory that produces an
ensemble identical to $\cal R$ can give falsifiable predictions. 
Genuine falsifiable prediction can only be made by a theory
whose ensemble $\cal M$ differs from $\cal R$. Further, to give
a genuine prediction there must be 
a property $\cal B$, not yet observed, but observable with 
present technology, which is probable in $\cal LM$ but improbable in $\cal LR$.  

A very important consequence of this follows from the following 
physical observation. 

{\it  Properties of the ensemble $\cal M$ generated
by a mechanism in a theory $\cal T$ will be random if that property
concerns physics on a scale many orders of magnitude from the scale
of the mechanism of production of universes defined by $\cal T$.}

One reason is familiar from statistical mechanics. Ensembles tend to
be randomized in observables that are not controlled in their 
definition. For example, for a gas in a room, the properties of 
individual atoms are randomized, subject only to their random values
being related to the temperature and density in the room.  

A second reason has to do with a general property of local field 
theories, which is decoupling of scales. In renormalizable field
theories, including those of the standard model, there is only weak
coupling between modes of the field at very different scales. 

We can see this applied to eternal inflation models. The mechanism of 
generation of universes involves quantum fluctuations in the presence
of a vacuum condensate with energy between $10^{15}$ and
$10^{20}$ Gev. Properties of the vacuum that influence physics at 
those scales will play a role in determining the ensemble of universes
created. If we consider the space of possible theories (perhaps string
vacua) theories will be preferentially selected by properties that
strongly influence the probability for a quantum fluctuation in this
environment to be uniform.  These will include coupling constants for
interactions manifest on that scale and vacuum expectation values for
Higgs fields also on that scale. But the exact values of masses or 
couplings of particles many orders of magnitude lighter are not going 
to show up. What will matter is the total number of degrees of 
freedom, but all particles so far observed are many orders of
magnitude lighter and may be treated as massless from the point of view 
of physics at the scale of universe creation. Hence, changes in the 
proton-neutron mass difference, or the electron-proton mass ratio
are not going to have a significant influence on the probability for
universe creation. The result is that these properties will be
randomized in the ensemble $\cal M$ created by eternal inflation. 

As a result, it is reasonable to expect that any  standard model parameters that govern low energy
physics, but do not govern physics at grand unified scales will have
the same distribution in $\cal M$ as in the random ensemble $\cal R$.
These include the masses of the quarks, leptons, and neutrino and the scale of 
electroweak symmetry breaking (and hence of the weak interactions).
{\it It follows that eternal inflation will not be able to make any 
falsifiable predictions regarding any of the low energy parameters parameters
of the standard model of particle physics. 
Consequently, no solution to the complexity problem can come from
eternal inflation, since that has to do with the values of these 
parameters. }

Eternal inflation may be able to make some predictions, but only those
restricted to parameters that govern physics at grand unified scales. 

There are claims that eternal inflation does lead, in  conjunction with another principle,
to predictions about the cosmological constant\cite{GV1,GV2,GV3,weinberg1,moreweinberg}. We next turn to an
examination of those claims. 

\subsubsection{The principle of mediocrity}

A variant of selection effects applied to a multi-universe is
the {\it the mediocrity principle}. This is 
defined by Garriga and Vilenkin\cite{GV1}
as requiring that  ``...our civilization is typical in the ensemble of 
all civilizations in the universe."

This means that we weigh the ensemble $\cal M$ by the number of
civilizations in each universe.  It follows that all universes
outside of $\cal LM$  have zero weight, and 
universes with more civilizations are weighed
more heavily.

Of course, it goes without saying that this principle adds several 
layers of presently untestable assumptions to the analysis. We know
nothing reliable about the conditions that generate civilizations. 
While we can speculate, our genuine knowledge about this is unlikely 
to improve in the near future. 

We can of course conjecture that the number of civilizations will be
proportional to the number of spiral galaxies. So we can provisionally
take the principle of mediocrity to mean that we weigh our ensemble
with the number of spiral galaxies in each universe.  Alternatively, we 
can postulate that the number of civilizations is proportional to the fraction of
baryons that end up in galaxies\cite{weinberg-recent}.  

Garriga and Vilenkin then argue that certain predictions can be drawn
concerning properties of the vacuum energy\cite{GV1}. We can note that,
in conformity with the above argument, no predictions are drawn 
concerning properties that are 1) have to do with the parameters
of low energy physics and 2) are uncorrelated in a random ensemble
with the existence of life. Still, it is good that people put 
predictions on the table and we should take them seriously.

To take them seriously we must ask, what exactly would be falsified
if one or more of their predictions is found to disagree with 
observation?  The argument depends on properties of the eternal 
inflation theory, some rough guesses about the wave function of the 
universe and how to reason with it, and some reasoning about the
effects of vacuum energy on the creation and evolution of galaxies,
as well as on their principle of mediocrity.  

The principle of mediocrity can only have force if it is more
stable than the other parts of the argument leading to the predictions. 
This must be so, otherwise a falsification of the prediction may
teach us only that the principle of mediocrity is unreliable. 
To be useful, a methodological principle  must be reliable enough
that it can be taken as firm, and used as part of an argument
to disconfirm an hypothesis about physics. 

So, is the principle of mediocrity on firmer ground than
quantum cosmology or the theory of galaxy formation?
What are the independent grounds for believing it?

I know of no {\it a priori} argument for the principle of mediocrity.
It may be the case that, if the multiverse is real, we live in
a universe with a maximal number of civilizations. But it could
just as easily be false. There is no reason why we may not live
in a universe which is untypical, in that it has some civilizations,
but many fewer than other members of the ensemble.  

Thus, while we can argue logically for taking into account selection
effects coming from the fact that we are in a universe hospitable to
living observers, the principle of mediocrity of \cite{GV1} is on much
less firm ground. 

The principle of mediocrity is sometimes supported by referring
to ensembles within which we as individuals are typical. And indeed,
there are many ensembles within which we as individuals are typical.
The problem is that there are also  many ensembles with respect to which
we are untypical. The principle of
mediocrity has little force in human affairs, because without further
specification it is vacuous, as we are both typical and atypical, 
depending on what ensembles we are compared against.

To see this, let us ask some questions about how typical we are.

\begin{enumerate}
    
    \item{} Do I live in the universe with the largest number
    of civilizations?
    
    \item{} Do I live in the universe with the largest number of
    intelligent beings?
    
    \item{} Do I live in the universe with the largest number
    of conscious minds?
    
    \item{} Do I live on the planet with the largest number of
    intelligent beings?
    
    \item{}Do I live in the most populous city on my planet?
    
    \item{}Do I live in the most populous country on my planet?
    
    \item{}Am I a member of the largest ethnic group on my planet?
    
    \item{}Do I have a typical level of wealth or income on my
    planet?
    
    \item{}Do I live at a time when more people are alive than
    at any other time?

\end{enumerate}    

The answers to questions {\bf 1-4} are that there is no way of telling, 
with either present data or future conceivable data. In my particular
case the answers to {\bf 5-8} are no, but I know people who can answer
yes to one or more of them. Question 
{\bf 9} is ambiguous. 
If the ensemble is all times in the past, the answer is 
probably yes. If the ensemble is all times, future and past, it is
impossible to know.

Given how often any individual fails to be typical in ensembles we 
know about, it seems to me we are on equally weak ground reasoning
from any assertion of answers to {\bf 1-4}  as any one of us would
be reasoning from {\bf 5-9}. 

The conclusion I draw is that
the principle of mediocrity is too ambiguous to be useful. It must
be supplemented by a specification of the ensemble. When that is done 
we can test it, but when we do we see that it is unreliable. 
Thus, it must be even less
reliable in situations where it cannot be tested.

Here is an example that illustrates the perils of the use of the principle of mediocrity.
This is a well known argument, called the {\it doomsday argument\cite{doomsday}.}  Someone begins
it by stating, {\it I am a typical human being.}  They may support that by noting the existence
of some ensembles within which they are typical. Then they introduce a new ensemble,
$\cal H$ consisting of {\it all human beings who will ever live.} They then assert that
since they are generally typical they should be typical in that ensemble. They
then proceed to draw a drastic deduction from this, we will call $\cal C$. This is that
{\it roughly the same number
of human beings will live after them as lived before them. }

Given that the population has
been growing exponentially for a long time, this leads to the conclusion that the
population should begin to fall drastically within their lifetime. 

There are more details, but we do not need them to see the ways in which the
argument is fallacious\footnote{Another criticism of the argument, from F. Markopoulou, is
that to even state that a person is typical in the ensemble $\cal H$ with respect to a
given property, is to assume that there is a normalizable probability distribution for
that property in $\cal H$. If the property is birth order, then the normalizability of the 
probability distribution already implies  
the population must decrease at some point in the future.   
Thus, the argument assumes 
what it claims to demonstrate\cite{fotini-personal}. The only thing left open is when, but as we see, this
cannot in any case be determined by the argument.}.  
The ensemble $\cal H$ contains an unknown number
of human beings, who may live in the future.  There is no way, given any information
we have at present, to determine if we, living now, are in any way typical or
untypical members of $\cal H$.  There is simply no point in guessing\footnote{It would take us
too far afield to analyze why such a fallacious argument is so attractive.  It has something to do with
the fallacy that every statement that will, at the end of time, have a truth value,  has a truth value
now.  The statement ``I  am a typical member of the ensemble $\cal H$" is one that can
only be given a truth value by  someone  in the unhappy situation of 
knowing they are the last of us, and they would thus judge it false.   No one for whom the statement
is true could possibly have enough information to ascribe to it a truth value, for the simple
reason that to do so would require knowledge of the future.  
The point can be put simply by saying that the logic of
truth values that can be ascribed by humans to questions about themselves is Heyting rather
than Boolean\cite{inside}.}.  Whether we who
have lived so far constitute most of $\cal H$, an infinitesimal fraction of $\cal H$ or
something in between depends on events that will take place in the future, most
of whom we are unable to control, let alone predict.  So it is simply impossible
on current knowledge to deduce the truth value of $\cal C$. 

However we can do one thing. We can look to the past. The population has
been growing exponentially for at least ten thousand years. Any person living in the
last ten thousand years would have had exactly as much rational basis to make
the argument starting with {\it "I am a typical human being"} and ending with 
the conclusion  $\cal C$ as we have-no more and no less. Other facts such as the
existence of weapons of mass destruction or global warming are irrelevant,
as they are not used to support $\cal C$. The whole point of the argument is 
supposed to be that
it is independent of facts such as these.

Now, we can ask, would a person have been correct to use this argument to conclude $\cal C$ 
a thousand years  in the past. NO-we know that they would have been wrong, because
already many more people have lived since them than lived before them.
But $\cal C$ is supposed to be a consequence of the mediocrity principle.
The conclusion is that  there are two cases of individuals to which the
principle may be applied. There is a class of individuals to whom the truth
value of $\cal C$-and hence of the mediocrity principle-cannot be checked.
Then there is a class of individuals about whom the truth value of $\cal C$ can
be determined. In each and every one of these cases, $\cal C$ is false.
Thus, in every case in which there is an independent check of the
consequences of the mediocrity principle, it turns out to be false. 
Hence, it is either false, or undetermined. Hence there is no evidence for
its truth. 

\subsubsection{Weinberg's argument for the cosmological constant}

Recently it has been claimed that the Anthropic Principle, and more specifically the Principle
of Mediocrity, led to a successful prediction.  This was Weinberg's prediction for the 
value of the cosmological constant, first made in \cite{weinberg1}, and then elaborated
in \cite{moreweinberg,weinberg-recent}.   It is important to note that Weinberg and collaborators 
actually make two separate arguments. The first  is the following: assuming {\bf A} and {\bf B} then we cannot find
ourselves in a universe with too large a positive value of $\Lambda$, or galaxies would never have
formed.  The upper limit for $\Lambda$, with all other constants of nature fixed, predicted by this
argument is about 200 times the present matter density\cite{weinberg-recent} (baryons plus
dark matter)  which 
gives roughly $\Omega_\Lambda < 100$.  This is about 
two orders of magnitude larger than the present observed value. 

In their second argument, Weinberg
and collaborators attempt to improve this estimate by evoking the principle of mediocrity in the form just discussed.  
They find that the 
probability to find a cosmological constant giving $\Omega_\Lambda$ of  $.7$ or less is
either $5 \%$ or $12\% $ depending on technical assumptions made.  Thus, one can conclude that while the
actual observed value is a bit low, compared to the mean, it is not at all unreasonable to argue that the
present observed value is consistent with the result of the analysis based on the Principle of Mediocrity.

Is there then something wrong with the case I've just presented against the Principle of Mediocrity? Someone
might then argue that, given that Weinberg's first paper came before the supernova and other recent
CMB measurements giving a value to $\Lambda$, his use of the Principle of Mediocrity has to count as a 
successful prediction. Indeed, this is perhaps the only successful new prediction in fundamental 
physics for the value of a physical parameter in decades\footnote{It is sometimes stated that Weinberg
made the only correct prediction for the order of magnitude of the cosmological constant. This is not true. A correct
prediction was made by Sorkin\cite{sorkin-cosmo}, based on the causal set approach to quantum gravity.  }.   

To reply, let us first distinguish the two arguments from each other.  One can reasonably conclude that
Weinberg's first argument is in part correct.  Were $\Omega_\Lambda > 100$ (with all other constants of nature fixed) 
we would have a problem understanding
why we live in a universe filled with galaxies.   However, this is an argument of the form we discussed above
in section 5.1.3-4: it is a false use of the Anthropic Principle. For the problem has nothing to do with our own
existence.  Just as Hoyle's argument has nothing to do with life, but is only based on the observed fact
that carbon is plentiful, Weinberg's first argument has to do only with the observed fact that galaxies are plentiful.
The argument itself can be made by a robot, or by a disembodied spirit. Were $ \Omega_\Lambda > 100$ (with all
other constants fixed) there
would be a contradiction between present models of structure formation in the universe, which are based on
established physical principles, and the observed fact that the universe is filled with galaxies.  

The first argument of Weinberg is sometimes presented as an example of the success of a selection
principle within a multiverse. But it then follows exactly the schema given at the beginning of 5.1.4, with
$X$ now the existence of many galaxies and $Z$ that $\Omega_\Lambda < 100$.  Were $\Omega_\Lambda$ in
fact equal to $1000$, the problem could not be with the multiverse hypothesis. It would be rather to explain
how galaxy formation happened despite such a large $\Lambda$, {\it  in a single universe.}  Therefore, Weinberg's
first  argument
is partly valid, but the part that is valid is just a rational deduction from an observed fact, that there are
galaxies. The mentions of the existence of life and selection effects in 
a multiverse are completely irrelevant to the actual logical structure of the argument, as they can be
removed from the argument without its logical force being in any way diminished.  

Before considering the second argument,  there is a caveat to deal with, which is the restriction to 
considering a class of universes in which all the other constants of nature are fixed.  As pointed out by
Rees\cite{Rees},  Tegmark and Rees\cite{Tegmark-Rees} and 
Graesser, Hsu, Jenkins, and  Wise\cite{wise-etal} it is difficult to justify  any claim to   
make a valid prediction based on this restricted assumption.  One should instead consider ensembles
in which other cosmological parameters are allowed to vary. When one does this the constraint on
$\Lambda$ from Weinberg's first argument is considerably weakened. They show this by considering
the case of the magnitude of the density fluctuations, usually denoted $Q$, which is observed to be about $10^{-5}$.  
One can argue that holding $\Lambda$ fixed, $Q$ cannot be much more than an order of magnitude
larger, for similar reasons. But, as they show, one can have stars and galaxies in a universe
in which {\it both } $Q$ and $\Lambda$ are raised several orders of magnitude from their present value
(see Figure 2 of \cite{Rees}).  

Thus, to illustrate the point just made, were $\Omega_\Lambda > 100$, one option to be considered
would be to raise $Q$, so as to explain the observed existence of galaxies in our universe. 

From the calculations of \cite{Rees,Tegmark-Rees,wise-etal}, we have to conclude that, if  
we make no other assumptions,  the pair $Q,\Lambda$ are each about
two orders of magnitude smaller than their most likely values, based only on reasoning from
the existence of stars and galaxies.  This is non-trivial, after all, they are many orders of magnitude
away from their natural values, in Planck units.  But it still leaves a great deal to be explained.

Let us now turn to Weinberg's second argument, in which he employs the Principle of Mediocrity.
In the last subsection I argued that this Principle provides an unreliable basis for deductions, because
the results gotten depend strongly on which ensemble one is considered to be a typical member. 
Does Weinberg's argument contradict or provide support for this conclusion? 
It is easy to see that it provides support.

Were Weinberg's second argument reliable, it would be 
robust under reasonable changes of the
ensemble considered.  But, as shown by Graesser et al \cite{wise-etal}, 
it is not\footnote{Related results were found earlier in \cite{GV2,GV3}.}.  If the ensemble is
taken to be universes in which $\Lambda$ varies, but all other constants are held fixed, then an application 
of the Principle of Mediocrity leads to the conclusion that 
the probability $\Lambda$ is as small as it is, is around $10 \%$.  But, if we consider an ensemble
in which $Q$ as well as $\Lambda$ varies, the probability comes down to order $10^{-4}$, with the precise
estimate depending
on various assumptions made  (see Table 1 of \cite{wise-etal}).  

We draw two conclusions from this. First, the Principle of Mediocrity is, as was argued, unreliable, because
the conclusions drawn from it are ambiguous in that they depend strongly on the ensemble considered. Second,
if, in spite of this, it is taken seriously, it leads to the conclusion that the probability for the observed
value of $\Lambda$ is on the order of one in ten-thousand. This follows from the fact  that, in all modern
cosmological theories, $Q$ depends on the parameters of the inflaton potential, such as its mass and
self-coupling.  In any fundamental model of particle physics, in which parameters vary, these would 
certainly be among the parameters expected to vary.  

Thus, the example of Weinberg's two arguments illustrates the conclusions we reached earlier.

The Anthropic Principle itself, in the form of {\bf A} and {\bf B} makes no predictions. Arguments that
it has led to predictions are fakes: what is actually doing the work in the arguments is never the existence 
of life or intelligent
observers, but only true observed facts about the universe, such as that carbon and galaxies are plentiful.

And we see here in detail that the Principle of Mediocrity cannot help, because it is easily shown to be
unreliable. Any argument that it leads to a correct conclusion can be easily turned into an argument for 
an incorrect conclusion by reasonable changes in the definition of the ensemble in which we are
assumed to be typical.  

\subsubsection{Aguirre's argument against the Anthropic Principle}

Before moving on to look at an example of a falsifiable theory, we mention one final argument
against the Anthropic Principle, given by Aguirre in \cite{Aguirre}.  He simply points out that intelligent
life would be possible in universes with parameters chosen very different from our own.  He gives
a particular example, which is a cold big bang model.  This is a class of examples, which were studied
sufficiently to show they disagree with present observations. Yet they still have galaxies, carbon chemistry 
 and long lived stars. 
This is sufficient, because it follows that any
argument that incorporates a version of the Anthropic Principle would have to explain why we don't
live in a cold big bang universe.  Given that cold big bang universes share the property of  our universe 
of having abundant formation of galaxies and stars, none of the versions of the Anthropic Principle can
do this. Thus, either we leave unexplained why we don't live in a cold big bang universe or we have
to find another explanation other than the Anthropic Principle for the parameters of our universe. 

\subsection{Natural Selection}

I believe I have said enough to demonstrate conclusively  that  the version of the Anthropic Principle 
described by {\bf A} and {\bf B}  is never going to give falsifiable predictions for the 
parameters of physics and cosmology.  The few times an argument called anthropic
has led to a successful prediction, as in the case of Hoyle's argument and Weinberg's
first argument, examination shows that the actual logical argument makes no reference
to an anthropic principle, but instead rests entirely on a straightforward deduction from
an observed fact about our universe.   

Thus,  if we are to understand the choices of parameters in 
detail, in the context of a falsifiable theory, we need an 
alternative approach. One alternative to 
deriving predictions from a multiverse
theories is patterned on the successful model of natural selection in
biology.  This was originally motivated by asking the question, where 
in science is there a successful solution to a problem of explaining 
improbable complexity? To my knowledge, only in biology do we 
successfully explain why some parameters-in this case the genes of
all the species in the biosphere-come 
to be set to very improbable values, with the consequence that the
system is vastly more complex and stable than would be for random
values. The intension is then not to indulge in some mysticism about
``living universes", but merely to borrow a successful methodology
from the only area of science that has successfully solved a problem
similar to the one we face\footnote{Other approaches to cosmology which employ
phenomena analogous to biological
evolution have been proposed, including
Davies\cite{davies}, Gribbin\cite{gribbin}, Kauffman\cite{stu},
Nambu\cite{nambu}, Schweber, 
Thirring and Wheeler\cite{old}.  
We note that Linde sometimes employs the term ``Darwinian"
to describe eternal inflation\cite{andrei,moreandrei}
However, because each universe in eternal inflation has the
same ancestor, there is not inheritance and modification of parameters
analogous to the case of biology.} 

The methodology of natural selection, applied to multiverse theories,
is described by three hypotheses:

\begin{enumerate} 

\item{}A physical process produces a multiverse with long chains 
of descendents. 

\item{}Let $\cal P$ be the space of dimensionless parameters of
the standard models of physics and cosmology, and let the parameters
be denoted by $p$. 
There is a fitness function $F(p)$ on $\cal P$ which is equal to
the average number of descendents of a universe with parameters
$p$.  

\item{}The dimensionless 
parameters $p_{new}$ of each new universe differ, on average 
by a {\it small }
random change from those of its immediate ancestor. 
 Small here means with small with 
respect to the change that would be required to significantly change
$F(p)$.

\end{enumerate}

Their conjunction leads to a predictive theory, because, using
standard arguments from population biology, after many
iterations from a large set of random starts, the population
of universes, given by a distribution $\rho (p)$, is peaked around
local extrema of $F(p)$.  With more detailed assumptions more
can be deduced, but this is sufficient to lead to observational
tests of these hypothesis, because this implies the
prediction that: 
 
\begin{itemize}
\item{}$\cal S$:
{\it If $p$ is changed from the present value in any direction in
$\cal P$ the first significant changes in $F(p)$ encountered
must be to decrease $F(p)$.}   

\end{itemize}

The point is that the process defined by the three hypotheses drives
the probability distribution $\rho (p)$ to the local maxima of the
fitness function and keeps it there. This is much more predictive
than the anthropic principle, because that principle resulting
probability distribution is much more structured, and very far from 
random. If in addition, the physics that determines the fitness 
function is well understood, detailed tests of the general
prediction $\cal S$ become possible, as we will now see.

\section{Predictions of Cosmological Natural Selection}

It is important to emphasize that the process of natural selection is
very different from a random sprinkling of universes on the 
parameter space $\cal P$. This would produce only a uniform 
distribution $\rho_{random} (p)$. To achieve a distribution peaked
around the local maxima of a fitness function requires the two 
conditions specified. The change in each generation must be small
so that the distribution can `climb the hills" in $F(p)$ rather
than jump around randomly, and so it can stay in the small volumes of
$\cal P$ where $F(p)$ is large, and not diffuse away. This requires
many steps to reach local maxima from random starts, which implies
that long chains of descendents are needed. 

As a result, of the two mechanisms of universe production so far 
studied, only black hole bouncing fits the conditions necessary
for natural selection to be applied. This is also fortunate, because
the physics that goes into the fitness function is, in this case,
well understood, at least in the neighborhood of the parameters
of our universe. The physical processes that strongly influence the number
of black holes produced are nucleosynthesis, galaxy formation,
star formation, stellar dynamics, 
supernova explosions, and the formation and stability 
of neutron stars. All of these stages, except, perhaps,  galaxy formation, are
understood in some detail, and in several of these cases our theories
make precise predictions which have been tested. We are then on
reasonably firm grounds asking what happens to each of these
processes when we make small changes in the parameters from their
present values. 

Thus, for the rest of this paper, by cosmological natural selection I
will mean the process of reproduction of universes through black hole
bounces, supplemented by the hypotheses above. 

The hypothesis that
the parameters $p$ change, on average by small random amounts,
should be ultimately grounded in fundamental physics. We note 
that this is compatible with string theory, in 
the sense that there are a great many string vacua, which likely
populate the space of low energy parameters well. It is plausible that
when a region of the universe is squeezed to Planck densities and 
heated to Planck temperatures, phase transitions may occur leading to
a transition from one string vacua to another. But there have so
far been no detailed studies of these processes which would check
the hypothesis that the change in each generation is small.  

One study of a bouncing cosmology, in quantum gravity, also lends 
support to the hypothesis that the parameters change in each 
bounce\cite{jorge-rodolfo-bounce}.

\subsection{Successes of the theory}

Details of tests of cosmological natural selection are described 
in \cite{evolution,moreev,lotc}.  I will only here summarize the conclusions here.
For details of the arguments, as well as references to the astrophysical literature
on which the arguments are founded, see \cite{evolution,moreev,lotc}.  

The crucial conditions necessary for forming many black holes as the
result of massive star formation are,

\begin{enumerate}
    
    \item{}There should be at least a few light stable nuclei, up to 
helium at least, so that gravitational 
    collapse leads to long lived, stable stars. 
    
    \item{}Carbon and oxygen nuclei should be stable, so that giant 
    molecular clouds form and cool efficiently, giving rise to the 
    efficient formation of stars massive enough to give rise to black 
    holes.  
    
    \item{}The number of massive stars is increased by feedback 
    processes by which massive star formation catalyzes more massive
    star formation. This is called ``self-propagated star formation,
    and there is good evidence that it makes a significant contribution
    to the number of massive stars produced. This requires a 
    separation of time scales between the time scale required for star 
    formation and the lifetime of the massive stars. This requires, 
    among other things, that 
     nucleosynthesis should not proceed too far, so that the universe 
    is dominated by long lived hydrogen burning stars.  
    
    \item{}Feedback processes involved in star formation also require 
    that supernovas should eject enough energy and material to 
    catalyze formation of massive stars, but not so much that there
    are not many supernova remnants over the upper mass limit for 
    stable neutron stars. 
    
    \item{}The parameters governing nuclear physics should be tuned, 
    as much as possible consistent with the forgoing, so that the 
    upper mass limit of neutron stars is as low as possible.

\end{enumerate}    

The study of conditions 1) to 4) leads to the conclusion that the number 
of black holes produced in galaxies will be decreased by any of the
following changes in the low energy parameters:

\begin{itemize}

\item{} A reversal of
the sign of  $\Delta m= m_{neutron} -m_{proton}$. 

\item{}A small  increase in $\Delta m$ (compared to 
$m_{neutron}$ will  destabilize 
helium and carbon.

\item{}An increase in $m_{electron}$ of order $m_{electron}$ itself,
will  destabilize 
helium and carbon.

\item{}An increase in $m_{neutrino}$ of order $m_{electron}$ itself,
will  destabilize 
helium and carbon.

\item{}A small increase in $\alpha$ will destabilize all nuclei.

\item{}A small decrease in $\alpha_{strong}$, the strong coupling 
constant, will destabilize all nuclei. 

\item{}An increase {\it or} decrease in $G_{Fermi}$ of order unity will 
decrease the energy output of supernovas. One sign will lead to a 
universe dominated by helium. 

\end{itemize}

Thus, the hypothesis of cosmological natural selection explains the
values of all the parameters that determine low energy physics
and chemistry: the masses of the proton, neutron, electron and
neutrino and the strengths of the strong, weak and electromagnetic
interactions. 

However, explanation is different from prediction. These cannot be
considered independent predictions of the theory, because the 
existence of carbon and oxygen, plus long lived stars, are also 
conditions of our own existence. Hence, selection effects prevent
us from claiming these as unique predictions of the theory of
cosmological natural selection.

If the theory is to make falsifiable tests, it must involve changes of 
parameters that do not effect the conditions necessary for our own
existence.  There are such tests, and they will be described shortly.

Before discussing them, however, we should address several criticisms 
that have been made.

\subsection{Previous criticisms}

Several arguments were made that $\cal S$ is in
fact contradicted by present observation \cite{ellis,harrison,silk}.  
These were found to
depend either on confusions about the hypothesis itself or 
on too simple assumptions about star formation. For example,
it was  argued in \cite{ellis} that star formation would proceed to more
massive stars were the universe to consist only of
neutrons, because there would be no nuclear processes to impede
direct collapse to black holes. This kind of argument ignores
the fact that the  formation of stars massive enough to
become black holes requires efficient cooling of giant molecular
clouds. The cooling processes that appear to be dominant require
carbon and oxygen, both for formation of $CO$, whose vibrational
modes are the most efficient mechanism of cooling, and because
dust grains, consisting of carbon and ice provide efficient
shielding of star forming regions from starlight. But
even processes cooling molecular clouds to $5-20^{o}$ K are
not enough, formation of massive stars appears to require 
that the cores of the cold clouds are disturbed by shock waves,
which come from ionized regions around other massive stars and
supernova.  For these reasons, our universe appears to produce many
more black holes than would a universe consisting of just 
neutrons\footnote{For details, see the appendix
of \cite{lotc}, which addresses the objections 
published in \cite{ellis,silk} and elsewhere.}.

Vilenkin\cite{AV-personal} raised the following issue
concerning the cosmological constant, $\Lambda$. He notes that were
$\Lambda$ (or vacuum energy) raised from the present value,
galaxy formation would not have taken place at all. one can
also add that even at a slightly increased value, galaxy formation would have
been cut off, leading only to small galaxies, unable to sustain the 
process of self-propagated star formation that is apparently necessary
for copious formation of massive stars.  This of course counts as
a success of the theory. 

On the other hand, were $\Lambda$ smaller than its present value,
there might be somewhat increased massive star formation, due to
the fact that at the present time the large spiral galaxies are
continuing to accrete matter through several processes. These
include the accretion of intergalactic gas onto the disks of galaxies
and the possible flow of gas from large gaseous disks that the visible
spiral galaxies may be embedded in.  It is of course difficult to 
estimate exactly how much the mass of spiral galaxies would be 
increased by this process, but Vilenkin\cite{AV-personal} as
much as $10 -20$ percent.  

However, lowering the cosmological constant would also increase the
number of mergers of spiral galaxies, and the number of absorptions
of dwarf galaxies by spirals.  These mergers and absorptions are
believed to convert spiral galaxies to elliptical galaxies, by
destroying the stellar disk and heating the gas. The result is to
cut off the formation of massive stars, leaving much gas not
converted to stars.  

There is then a competition between two effects. Raise $\Lambda$
and galaxies do not form, or do not grow large enough to support
disks and hence massive star formation. Decrease $\Lambda$ and the
dominant effect may be to cut of massive star formation, due to
increased mergers and absorptions converting spiral to elliptical 
galaxies.  One can conjecture that the present value of $\Lambda$ maximizes
the formation of black holes,

Other claims 
have been made
that with present knowledge $\cal S$ is in fact not 
testable\cite{silk,rees}.  Below I will  show that these claims
are also false, by  explaining why a single observation of an
astrophysical object that very well might exist-a heavy neutron
star-
would refute
$\cal S$.    After this I describe two more kinds of observations
that may be made in the near future which could lead to 
refutations of $\cal S$.  These have to do with more accurate
observations of  the
spectrum of fluctuations in the cosmic microwave background
($CMB$) and the initial mass function for star formation in
the absence of carbon.

\subsection{Why a single heavy pulsar would refute $\cal S$.}

Bethe and Brown, in \cite{bethebrown} introduced 
the hypothesis that neutron
star cores contain a condensate of $K^-$ mesons.  For the present
purposes their work can be expressed in the following way.
Calculations show\cite{bethebrown} that there is a 
critical value $\mu_c$
for the strange quark mass $\mu$ such that if $\mu < \mu_c$ then
neutron star cores consist of approximately equal numbers of
protons and neutrons with the charge balanced by a condensate
of $K^-$ mesons.  The reason is that in nuclear matter the effective
mass of the $K^-$ is renormalized downward by an
amount depending on the density $\rho$.   Given a choice of
the strange quark mass,
$\mu$, let  $\rho_0 (\mu )$ be the density 
where the renormalized Kaon mass is less than the electron mass.
$\mu_c$ is the value of $\mu$ where $\rho_0 (\mu )$ is less
than the density $\rho_e$ at which the electrons react with the
protons to form neutrons.  In either case one neutrino per 
electron is produced, leading to a supernova.  

Bethe, Brown and collaborators claim that calculations show
that $\mu <\mu_c$ \cite{bethebrown}.  But their calculations involve 
approximations
such as chiral dynamics and may be sufficiently inaccurate
that in fact $\mu_c > \mu$.  However, the
accuracy of the calculations 
increases as $\mu^{-2}$ as $\mu$ is decreased so, even
if we are not sure of the conclusion that $\mu < \mu_c$, we
can be reasonably sure of the existence of such a critical value
$\mu_c$.  Then we may reason as follows.  
If $\mu < \mu_c$ then, as shown by calculations\cite{bethebrown}
the upper
mass limit is low, approximately $1.5 M_\circ$.  If $\mu> \mu_c$
neutron stars have the conventional equations of state and the
upper mass limit is higher, almost certainly above $2$
\cite{uml}.  
Therefore a single observation of a neutron star whose mass
$M$ was sufficiently high would show that $\mu > \mu_c$,  
refuting Bethe and Brown's claim for the opposite.  
Sufficiently high is certainly $2.5 M_\circ$, although if
one is completely confident of Bethe and Brown's upper limit
of $1.5$ solar masses, any value higher than this would be
troubling.
Furthermore,
this would refute $\cal S$ because it would then be the case that
a decrease of $\mu$ would lead to a world with a lower upper
mass limit for neutron stars, and therefore more black holes.

Presently all well measured neutron star masses are from
binary pulsar data and are all below 
$1.5 M_\circ$ \cite{nsm}\footnote{Other methods yield less
precise estimates\cite{estimates}.}. But an 
observation of a heavy neutron star may be made at any time.  

We may note that this argument is independent of any issue
of selection effects associated with ``anthropic reasoning",
because the value of the strange quark mass $\mu$ may
be varied within a large range before it produces a significant
effect on the chemistry\footnote{Skeptics might reply that
were $\cal S$ so refuted it could be modified to a new
${\cal S}^\prime$,  which was not refuted by the addition of the
hypothesis that $\mu$ is not an independent parameter and cannot
be varied without also, say, changing the proton-neutron mass
difference, leading to large effects in star formation.  It is of course,
a standard observation of philosophers of science that most 
scientific hypotheses can be saved from refutation by the 
proliferation
of ad hoc hypotheses.  In spite of this, science proceeds by
rejecting hypotheses that are refuted in the absence of special
fixes.  There are occasions where such a fix is warranted.  The 
present case  would only be among them if there were a preferred
fundamental theory, such as string theory, which had strong
independent experimental support, in which it turned
out that $\mu$ was in fact not an independent parameter, but
could not be changed without altering the values of parameters
that strongly affect star formation and evolution. }.

\subsection{How observations of the $CMB$ could refute $\cal S$.}

It may be observed that the universe might have had many more
primordial black holes than it seems to have were the
spectrum of primordial fluctuations, $f(n)$ tilted 
to increase the
proportion on small scales\cite{silk}.  Of course, this observation by
itself does not refute $\cal S$ directly unless it is shown that
the standard model has a parameter that can be varied to
achieve the tilt in the spectrum.  It does not, but it is reasonable
to examine whether plausible extensions of the standard model
might.  One such plausible extension is to add fields that could
serve as an inflaton, so that the theory predicts inflation.  Given
an extension of the standard model, $\cal E$,  
that predicts inflation, the
spectrum of primordial fluctuations is in fact predicted as a 
function of the parameters of $\cal E$.  Thus, $\cal S$ is
refuted if a) some model $\cal E$ of inflation is observationally
confirmed and if b) that particular extension of the standard
model has some parameter, $p_{inf}$ that can be modified to
{\it increase the total numbers of black holes produced, including
primordial black holes.}  Given
the accuracy expected for observations of the $CMB$ from 
MAP and PLANCK, there is a realistic possibility that observations
will distinguish between different hypotheses for $\cal E$ and measure
the values of their parameters.

In the standard ``new" inflationary scenario\cite{newinflation} 
there is no
parameter that fulfills the function required of $p_{inf}$.   There
is the inflaton coupling $\lambda$, and it is true that the
amplitude of the $f(n)$ is proportional to $\lambda$ so that
the number of primordial black holes can be increased by
increasing $\lambda$.  However, the size of the region that
inflates $R$, is given by $R \approx e^{\lambda^{-1/2}}$.  This
effect overwhelms the possibility of making primordial black holes.
In fact, if the observations confirm that the standard new inflationary
scenario is correct, then $\cal S$ is refuted if $\lambda$ is not
tuned to the value that results in the largest total
production of black holes in the inflated region\cite{evolution}.
Because of the
exponential decrease in $R$ with increasing $\lambda$, this
is likely close to the {\it smallest} possible value that leads to a 
sizable constant density of black holes produced in comoving
volumes during the history of the universe.  This is likely
the smallest $\lambda$ that still allows prolific formation of
galaxies\cite{evolution}.

This seems consistent
with the actual situation in which there appears to have been
little production of primordial black holes, so that the primary mode
of production of black holes seems to be through massive star 
production, in galaxies that apparently do not form till rather late,
given that $Q= \delta \rho /\rho \approx 10^{-5}$.    

However, there are non-standard models of inflation that have
parameters $p_{inf}$ that can be varied from the present
values in a manner that tilts $f(n)$ so that more primordial
black holes are created than in our universe, without at the same
time decreasing $R$\cite{tiltinflation}.  If future observations 
of the $CMB$ cleanly
show that the standard new inflation is ruled out, and only models
with such a parameter $p_{inf}$ are allowed, then 
$\cal S$ will be refuted.  

This is a weaker argument than the first one,  
but given the scope for increases in
the accuracy of measurements of the $CMB$, and hence of
tests of inflationary models, it is a realistic possibility that
$\cal S$ may be refuted by such an argument.

\subsection{How early star formation could refute $\cal S$.}

As shown in \cite{carrees,rees,evolution} there 
are several directions in $\cal P$
which lead to universes that contain no stable nuclear
bound states.  It is argued  in \cite{evolution,lotc} this 
leads to a strong
decrease in $F(p)$ because the gravitational collapse of objects
more massive than the upper mass limit of neutron stars
in our universe seems to depend on the cooling mechanisms
in giant molecular clouds, which are dominated by 
radiation from $CO$.  In a universe without nuclear bound
states the upper mass limit for stable collapsed objects is
unlikely to decrease dramatically (as the dominant factor
ensuring stability is fermi statistics)  while without cooling
from $CO$ collapsed objects whose ultimate size is above
the upper mass limit are likely to be less common.  

In the
absence of bound states the main cooling mechanism appears
to involve molecular hydrogen\cite{early}, but 
there are two reasons to
suppose this would not lead to plentiful collapse of massive
objects in a world with nuclear bound states.  The first is that
in such a world there would be no dust grains which appear to
be the primary catalysts for the binding of molecular hydrogen.
The second is that in any case molecular hydrogen is a  
less efficient coolant than $CO$ \cite{early}.  

This is also a weaker argument than the first, given present
uncertainties in our understanding of star formation processes, but
as that understanding is likely to become more precise in the
near future let us follow it.  Could this argument 
be refuted by any possible
observations?  In the present universe the collapse of massive
objects is dominated by processes that involve nuclear bound
states, but we have available a laboratory for the collapse of
objects in the absence of nuclear bound states, which is the
universe before enrichment with metals.  Indeed, we know
that there must have been collapse of massive objects during
that period as otherwise carbon, oxygen and other elements
would not have been produced in the first place.   But given
that $CO$ acts as a catalyst for formation of heavy elements,
and that the dust formed from heavy elements produced in stars
is also a catalyst for molecular binding, 
there is an instability whereas any chance formation of massive
objects leads in a few million years to both an enrichment of the
surrounding medium and the production of significant quantities
of dust, and these greatly increase the probability for
the formation of additional massive objects.  Hence, the initial
rate of formation of heavy objects in the absence of enrichment
{\it does not have to be very high} to explain how the universe
first became enriched.

This shows  that the fact that there was
some collapse of heavy objects before enrichment does not refute
the argument that the number of black holes produced in a universe
without nuclear bound states would be much less than at
present.  But while it thus doesn't refute $\cal S$, it doesn't
establish it either.  It is still consistent with present knowledge
that the production of massive objects in the absence of
heavy elements proceeds efficiently under the right conditions,
so that there may have in fact been a great deal of early star
formation uncatalyzed by any process involving heavy elements.

This could lead to a refutation of $\cal S$ because, in a world
without nuclear bound states, many more  massive collapsed
objects would become black holes than do in our universe, where the
collapse is delayed by stellar nucleosythesis.  

The question is then 
whether a combination of observation and theory
could disentangle the strong catalytic effects of heavy elements
leading to a strong positive feedback in massive star formation
from the initial rate of massive star formation without heavy
elements.  
Although models of star formation with and without heavy
elements are not sufficiently developed to distinguish the
two contributions to early star formation, it is 
likely that this will become possible  as our ability to model star
formation improves.
If so than it is also possible that future observations will be
able to measure enough information about early star formation
to distinguish the two effects.  If the conclusion is that the number of
black holes formed is greater in world without nuclear bound
states than in our own then $\cal S$ would be refuted.

\section{Conclusions}

This article has been written with the hope of contributing to a debate about the possible role of the Anthropic 
Principle in physics and cosmology.  Closing it I find myself even more puzzled than when I began as to
why the Anthropic Principle has such strong support by so many otherwise  good scientists.  Having carefully considered the
arguments, and engaged several proponents who I deeply respect in conversation and correspondence, I
come to the conclusions I have explained here. The logic seems to me incontrovertible, and it leads to the conclusion
that not only is the Anthropic Principle not science, its role may be negative. 
 To the extent that the Anthropic Principle is espoused to justify continued  interest in 
unfalsifiable theories, it may play a destructive role in the progress of science.  

If I am mistaken about any of this, I hope someone will set me straight. The main points of my argument have
been:  

\begin{itemize}

\item{} No theory can be a candidate for a physical theory that does not make falsifiable
predictions. To violate this maxim is to risk the development of a situation in which
the scientific community splits into groups divided by different unverifiable faiths, because
there is no  possibility of killing popular theories by rational argument from shared evidence.

\item{}The version of the Anthropic Principle described by {\bf A} and {\bf B} cannot lead to falsifiable theories. 

\item{}There are successful predictions claimed to have come from anthropic reasoning. Examination shows
they all are of two kinds.  1) Uncontroversial use of selection effects within one universe, as in the argument of 
Dicke. 2) Simple
deductions from observed facts, with the mention of life, or ensembles of universes playing no actual
role in the argument leading to the successful prediction, as in the arguments of Hoyle and Weinberg.  
There are no successful predictions claimed to be Anthropic that do not fall into these two classes. 
Hence all claims for the success of the Anthropic Principle are false.  

\item{}The Principle of Mediocrity is ambiguous, because a reasonable change in the definition of the ensemble in which
we or our civilization are taken to be typical  can often
turn an argument for a correct conclusion into an argument for an incorrect conclusion.  In specific applications
we find it is often unreliable.  When the claims made can be tested, as in the doomsday argument
applied to our ancestors,  it often leads to false conclusions. 

\item{}We also find that eternal inflation cannot in any case lead to an explanation of the low energy
parameters of the standard model, and thus to a resolution of the problem of why the parameters are chosen
so as to allow stars and organic chemistry, because the values of the low energy parameters play no role
in the mechanism that generates the probability distribution for universes created by eternal inflation.  

\item{}It is possible to derive falsifiable predictions from a multiverse theory, if the following conditions are
satisfied: 1) The ensemble of universes generated must differ strongly from a random ensemble, constructed
from an unbiased measure.  2) Almost all members of the ensemble must have a property $\cal W$ that
is not a consequence of either the known laws of physics or a requirement for the existence of life.  3) It must
be possible to establish whether $\cal W$ is true or not in our universe by a doable experiment.  

\item{}There is at least one example of a falsifiable theory satisfying these conditions, which is cosmological
natural selection.  Among the properties $\cal W$ that make the theory falsifiable is that the upper mass
limit of neutron stars is less than $1.6$ solar masses.  This and other predictions of CNS have yet
to be falsified, but they could easily be by observations in progress.

\end{itemize}

It  must then be considered unacceptable for any theory, claimed to be a fundamental
theory of physics, to rely  on the Anthropic Principle to make contact
with observations.  When such claims are made, as they have been recently for 
string theory\cite{Susskind} this can only be considered signs that a theory is in
deep trouble, and at great risk of venturing outside the bounds of science. 

There are of course alternatives. String theory could possibly be shown to imply the conditions
necessary for cosmological natural selection to be applied, in which case it would yield falsifiable
predictions.  Or another mechanism of selection of parameters could be found that does lead
to falsifiable predictions.  What is clear is that some falsifiable version of the theory must be 
found.   If not, 
the theory cannot be considered a scientific theory, because there will be no way to establish
its truth or falsity by a means which allows consensus to be established by rational argument from
shared evidence.  

\section*{ACKNOWLEDGEMENTS}

I am grateful to numerous people for conversation and correspondence on these topics, including
Antony Aguirre, Stephon Alexander, Tom Banks, John Barrow, Michael Dine, George Ellis, 
Ted Jacobson, Joao Magueijo,  Fotini Markopoulou, 
John Moffat, Slava Mukhanov, Holger  Nielsen, Joe Polchinski,  Martin Rees,  Sherrie Roush,  Alex Vilenkin, Peter Woit, 
and colleagues at Perimeter Institute.   
I am also grateful to Michael Douglas, Jaume Garriga,  Leonard Susskind, Alex Vilenkin 
and Steven Weinberg for helpful clarifications
and references.  
This research was supported partly by the Jesse Phillips 
Foundation. Finally, I am grateful to the 
Templeton Foundation and the organizers of the Cambridge Conference on the Anthropic Principle
for including anthro-skeptics like myself in the debate, in the true spirit of science.

\end{document}